\pdfoutput=1

\documentclass{appolb}

\usepackage{amssymb}
\usepackage{hyperref}
\usepackage{graphicx}
\usepackage{amsmath}
\usepackage[numbers,sectionbib,sort&compress]{natbib}

\usepackage{graphicx,amsmath}

\begin{document}
\title{Correlations in ultra-relativistic nuclear collisions with strings%
\thanks{Presented at Excited QCD 2019, Schladming, Austria; Supported by Polish National Science Center grant 2015/19/B/ST/00937}
}
\author{Martin Rohrmoser$^a$\thanks{Martin.Rohrmoser@ujk.edu.pl}, Wojciech Broniowski$^{a,b}$
\address{$^a$ Institute of Physics, Jan Kochanowski University, PL-25406 Kielce, Poland\\
$^b$ The H. Niewodnicza\'nski Institute of Nuclear Physics,
Polish Academy of Sciences, PL-31342 Cracow, Poland
}
}

\maketitle
\begin{abstract}
While string models describe initial state radiation in ultra-relativistic nuclear collisions well, they mainly differ in their end-point positions of the strings in spatial rapidity. We present a generic model where wounded constituents are amended with strings whose both end-point positions fluctuate and analyze semi-analytically various scenarios of string-end-point fluctuations.
In particular we constrain the different cases to experimental data on rapidity spectra from collisions at $\sqrt{s_{\rm NN}}=200$~GeV, and explore their respective two-body correlations, which allows to partially discriminate the possible solutions.

\end{abstract}

Our main goal is to better understand the origin of forward-backward multiplicity fluctuations within ultrarelativistic nuclear collisions.
This text is mainly based on our work~\cite{Rohrmoser:2018shp}, which generalizes the analysis of~\cite{Broniowski:2015oif}.
Our approach uses strings with fluctuating end-points together with fluctuations in the number of sources of these strings in order to describe the multiplicity fluctuations.

QCD-motivated string models have been successful in describing soft particle production  
-- in particular Monte-Carlo implementations of the Lund model~\cite{Andersson:1983ia,Wang:1991hta,Lin:2004en,Sjostrand:2014zea,Bierlich:2018xfw,Ferreres-Sole:2018vgo} or the Dual Parton model involving Pomeron and Regge exchange~\cite{Capella:1992yb,Werner:2010aa,Pierog:2013ria}. 
These models have in common that they assume the formation of numerous strings at early stages of nuclear collisions. 
These strings represent the confined color fields spanned between two opposite color charges. 
Breakings of these strings correspond to particle-antiparticle creation and accounts for the large multiplicity creation at the early stages of nuclear collisions.
However, distributions of string-end points vary between the different approaches. 
Thus, we also try to understand the phenomenological consequences of different string-end-point distributions.

On the other hand, the produced multiplicity can be successfully  described within the wounded picture~\cite{Bialas:1976ed}. In particular the  wounded constituent model~\cite{Bialas:1977en,Bialas:1977xp,Bialas:1978ze,Anisovich:1977av} works remarkably well  
in the description of RHIC data. 
The wounded picture describes the spectra $\frac{dN}{d\eta}$ via the creation of a number of sources within the Glauber model~\cite{Glauber:1959aa} which all emit particles following a common emission profile $f(\eta)$.
 
Before merging the two models, we will outline the wounded constituent model, which we write as
\begin{eqnarray}
\frac{dN}{d\eta} = \langle{N_A}\rangle f(\eta)+\langle{N_B}\rangle f(-\eta), \label{eq:woundeta}
\end{eqnarray}
where ${N_A}$ and ${N_B}$ are the number of wounded constituents. 
For our numerical results these numbers were obtained by GLISSANDO~\cite{Rybczynski:2013yba} a Monte-Carlo simulation code of the Glauber model, where it was assumed that every nucleon can provide up to three wounded constituents.

\begin{figure}
\centering
\includegraphics[scale=0.55,trim=0 10 0 0, clip=true]{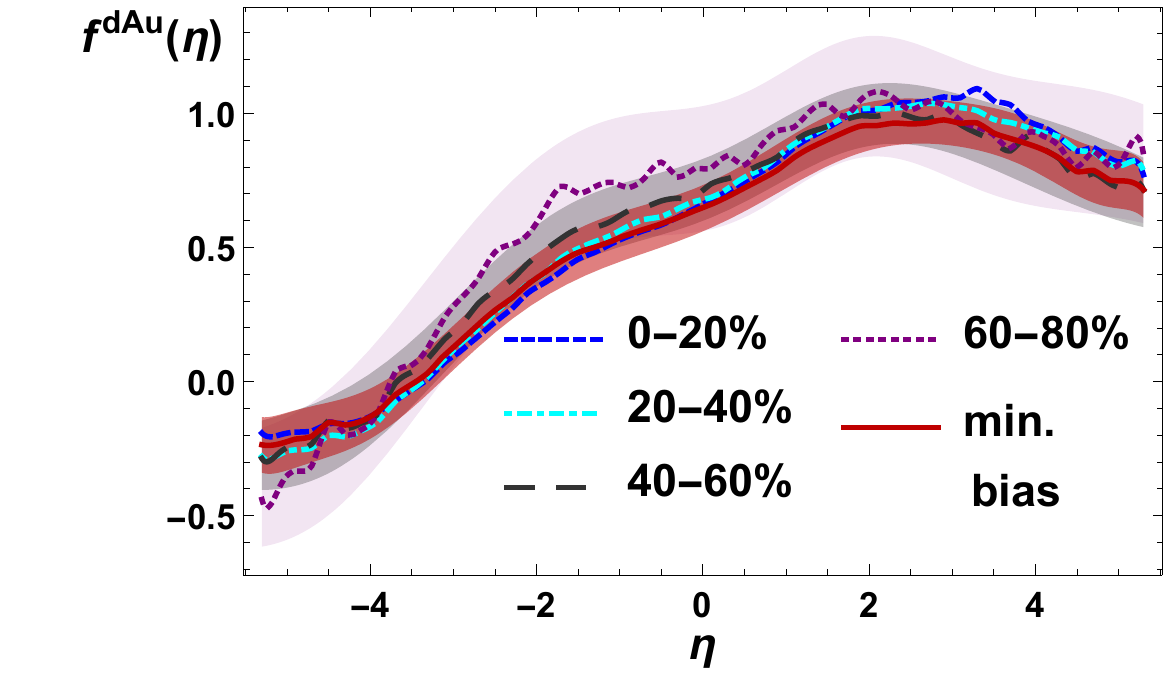}
\caption{Emission profiles from PHOBOS data~\cite{Back:2003hx,Back:2004mr} on d-Au collisions at $\sqrt{s_{\rm NN}}=200$~GeV. Experimental uncertainties have been propagated for the $40 - 60\%$ and $60 - 80\%$ centrality classes and the minimum bias case (colored bands).}
\label{fig:scaling}
\end{figure}

We verified the scaling behavior of Eq.~(\ref{eq:woundeta}) by extracting from experimental data an emission profile, which does not depend on the number of sources.
Fig.~\ref{fig:scaling} shows an example for experimental data from PHOBOS~\cite{Back:2003hx,Back:2004mr} on d-Au collisions.
As it can be seen the extracted emission profiles overlap within the uncertainties propagated from experiment, so that a description of rapidity spectra with a universal emission profile $f(\eta)$ can be justified. We thus confirm the results by~\cite{Barej:2017kcw}.

We also verified, whether it is possible to reproduce both spectra for d-Au and Au-Au collisions, comparing to PHOBOS data~\cite{Back:2003hx,Back:2004mr,Back:2002wb}. To this end, we used an emission profile $f(\eta)$ with a symmetric part obtained from Au-Au and and asymmetric part obtained from d-Au collisions and could reproduce the qualitative behavior of the spectra. We will use this particular version of  $f(\eta)$ in the remainder of the text.
 
The wounded constituent model is combined with a generic string model, where each wounded source pulls exactly one string in pseudo-rapidity with end-points $y_1$ and $y_2$. The strings are assumed to break at least once, which yields particle emission at pseudo-rapidity $\eta$, which follows for each string individually a radiation profile $s(\eta;y_1,y_2)$.
For simplicity we assume uniform probability for particle emission, i.e.:
\begin{eqnarray}
s(\eta;y_1,y_2)=\omega\left[\theta(y_1<\eta<y_2)+\theta(y_2<\eta<y_1)\right],
\end{eqnarray}
where $\omega$ is the production rate. 

We use string-end-point distributions $g_1(y_1)$ and $g_2(y_2)$ for which we demand that they allow to reproduce the one-body-emission profile $f(\eta)$ extracted from experiment. 
Thus, we find that
\begin{eqnarray}
f(\eta) =\int_{-\infty}^{\infty} \!\!\! dy_1\,g_1(y_1) \int_{-\infty}^{\infty}\!\!\! dy_2\, g_2(y_2) s(\eta,y_1,y_2)= 
 \omega \left [ \tfrac{1}{2} - 2 H_1(\eta)  H_2(\eta)  \right ], \label{eq:f1G} 
\end{eqnarray}
with the shifted cumulative distribution function defined as
\begin{eqnarray}
H_{i}(\eta)=G_{i}(\eta)-\tfrac{1}{2}, \;\;\; G_i(\eta)=\int_{-\infty}^\eta dy \, g_{i}(y), \;\; i=1,2,
\end{eqnarray}

\begin{figure}
\includegraphics[scale=0.49,clip=true,trim=0 10 0 0]{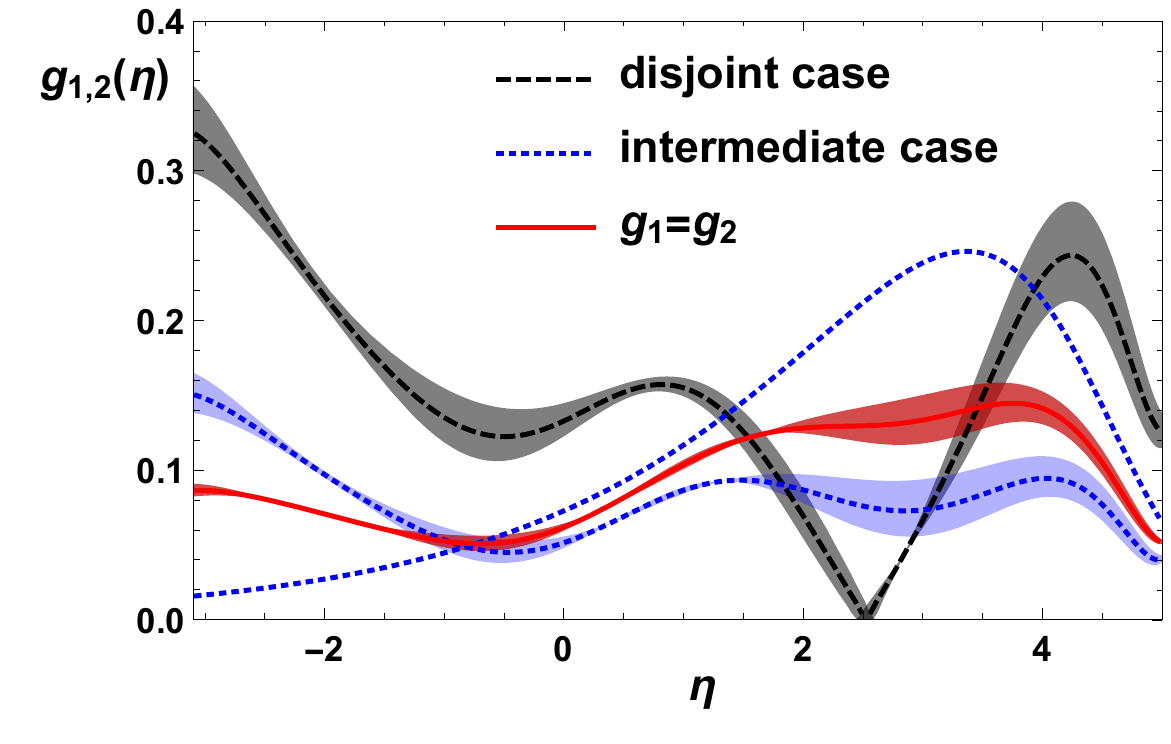}
\includegraphics[scale=0.49,clip=true,trim=0 10 0 0]{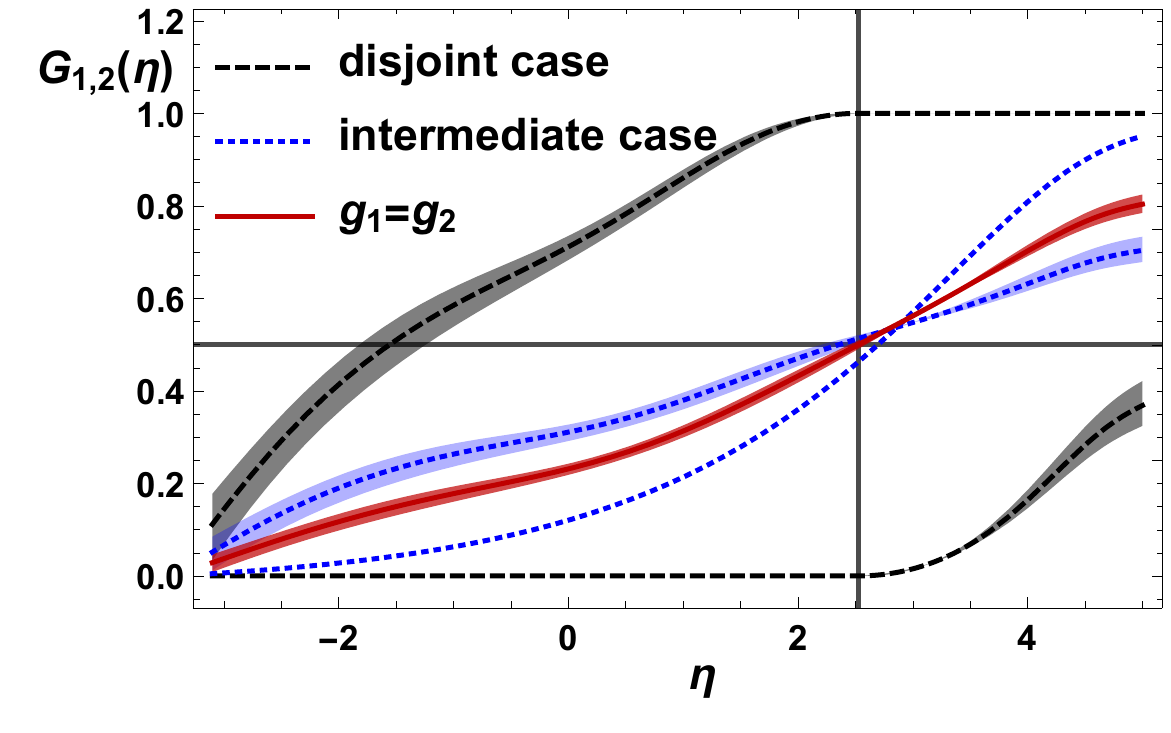}
\caption{The three solutions for string-end point distribution functions (left) discussed in the text together with their cumulative distribution functions (right). }
\label{fig:gi}
\end{figure}
It is clear from Eq.~(\ref{eq:f1G}) that choices for $H_1(\eta)$ and $H_2(\eta)$ are not unique.
However, since  $H_1(\eta),H_2(\eta)\in[-\frac{1}{2},\frac{1}{2}]$ one can infer  
$\omega\in[f(\eta_{\rm max}),2f(\eta_{\rm max})]$, where $\eta_{\rm max}$ is the position in pseudo-rapidity of the maximum of $f(\eta)$ . 
We study the following three cases of solutions to Eq.~(\ref{eq:f1G}):
\begin{enumerate}
\item $\omega=2f(\eta_{\rm max})$ (we label the case as ``$g_1=g_2$''), where one obtains
\begin{eqnarray}
H_1(\eta)=H_2(\eta)=\sqrt{\frac{1}{4}- \frac{1}{2\omega} f(\eta)}\,{\rm sgn}(\eta-\eta_{\rm max}). \label{eq:case1}
\end{eqnarray}
\item $\omega=f(\eta_{\rm max})$ (labeled as "disjoint case"), where one obtains
\begin{eqnarray}
H_1(\eta)&=& - \frac{1}{2} \theta(\eta_{\rm max}-\eta) +\left [ \frac{1}{2} - \frac{1}{\omega} f(\eta) \right ] \theta(\eta-\eta_{\rm max}), \nonumber \\
H_2(\eta)&=& - \left [ \frac{1}{2} - \frac{1}{\omega} f(\eta) \right ] \theta(\eta_{\rm max}-\eta)  + \frac{1}{2} \theta(\eta-\eta_{\rm max}).  \label{eq:case2}
\end{eqnarray}

\item an intermediate case, where $f(\eta_{\rm max})<\omega<2f(\eta_{\rm max})$. There, $H_1(\eta)$
is assumed as fixed and one obtains
\begin{eqnarray}
H_2(\eta)=\frac{\frac{1}{4}- \frac{1}{2\omega} f(\eta)}{H_1(\eta)}. \label{eq:h2}
\end{eqnarray}
\end{enumerate}
One can conclude from Eq.~(\ref{eq:f1G}) that the solutions for  $H_1(\eta)$ and $H_2(\eta)$ in the disjoint case serve as upper and lower limits for all other solutions to  Eq.~(\ref{eq:case2}).
Fig.~\ref{fig:gi} shows results for $g_1(\eta)$ and $g_2(\eta)$ as well as $G_1(\eta)$ and $G_2(\eta)$.
 
Eq.~(\ref{eq:f1G}) can be generalized to obtain the density $f_2(\eta_1,\eta_2)$ for the emission of particle pairs at pseudo-rapidities $\eta_1$ and $\eta_2$ as~\cite{Rohrmoser:2018shp}
\begin{eqnarray}
f_2(\eta_1,\eta_2) = \omega^2  \label{eq:fnG} G_1[{\rm min}(\eta_1,\eta_2)] \left \{1-G_2[{\rm max}(\eta_1,\eta_2)] \right \} + (1 \leftrightarrow 2).
\end{eqnarray}

\begin{figure}
\centering
\includegraphics[scale=0.49,clip=true, trim=0 0 0 0]{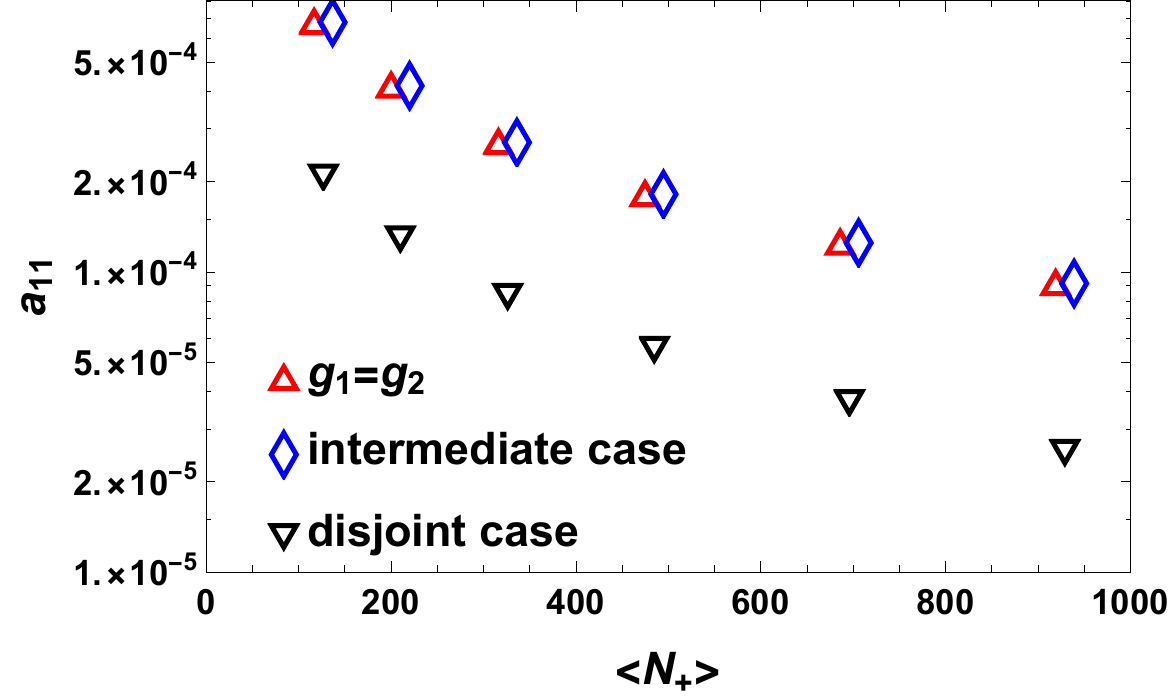}
\includegraphics[scale=0.47]{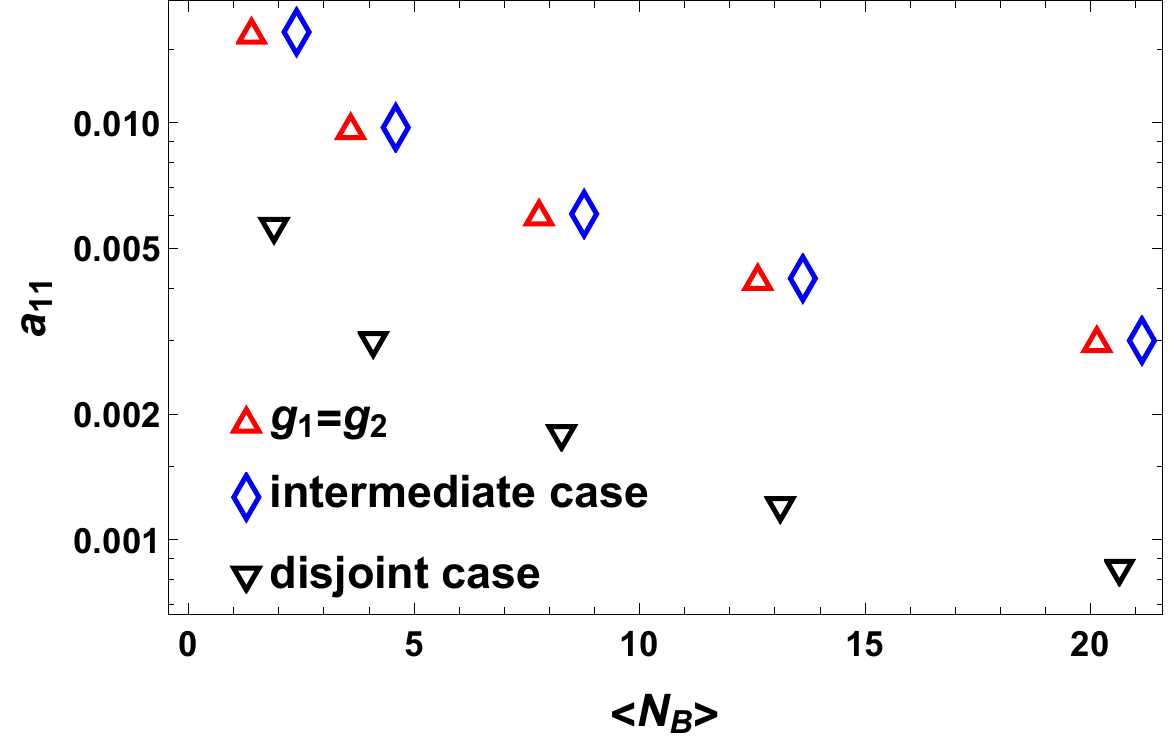}
\caption{Legendre coefficients $a_{11}$ for Au-Au (left) and d-Au collisions (right) at $\sqrt{s_{\rm NN}}=200$~GeV as a function of the number of sources, where $N_+=N_A+N_B$.}
\label{fig:a11}
\end{figure}

Summing over all possible sources one can obtain the covariance for the emission of particle pairs in nuclear collisions as 
\begin{eqnarray}
 {\rm cov}_{AB}(\eta_1, \eta_2)  
                            &=& \langle{N_A}\rangle {\rm cov}(\eta_1,\eta_2) + \langle{N_B}\rangle {\rm cov}(-\eta_1,-\eta_2)   \label{eq:gen} \\
                            &+& {\rm var}(N_A) {f(\eta_1)}{f(\eta_2)} + {\rm var}(N_B) {f(-\eta_1)}{f(-\eta_2)} \nonumber \\
                            &+& {\rm cov}(N_A,N_B) \left [{f(\eta_1)} {f(-\eta_2)}+ {f(-\eta_1)} {f(\eta_2)} \right ], \nonumber
\end{eqnarray}
with ${\rm cov}(\eta_1,\eta_2)={f_2(\eta_1,\eta_2)} -  {f(\eta_1)} {f(\eta_2)}$.
One can also define the correlations $C$ as
\begin{eqnarray}
C_{AB}(\eta_1,\eta_2) =1+ \frac{{\rm cov}_{AB}(\eta_1,\eta_2)}{f_{AB}(\eta_1)f_{AB}(\eta_2)},\qquad {\rm with} f_{AB}(\eta):=\frac{dN}{d\eta} \label{eq:C}
\end{eqnarray}
One can define $a_{nm}$  coefficients~\cite{Bzdak:2012tp} 
as projections of $C_{AB}(\eta_1,\eta_2)$ on $T_n(x)=\sqrt{n+1/2}P_n(x)$ (with Legendre polynomials $P_n(x)$ )
\begin{eqnarray}
a_{nm} &=& \frac{\int_{-Y}^Y {d \eta_1} \int_{-Y}^Y {d \eta_2} C_{AB}(\eta_1,\eta_2)
T_n\left(\frac{\eta_1}{Y}\right) T_m\left(\frac{\eta_2}{Y}\right)}
{\int_{-Y}^Y {d \eta_1} \int_{-Y}^Y {d \eta_2} C_{AB}(\eta_1,\eta_2)}, \label{eq:anmC}
\end{eqnarray}
with the covered pseudo-rapidity range $[-Y,Y]$, where we use $Y=1$ for RHIC.
Results for $a_{11}$ are shown in Fig.~\ref{fig:a11}. 
They scale to a good approximation as the inverse number of sources, as can be expected from Eqs.~(\ref{eq:gen})-(\ref{eq:anmC}).
Furthermore, the $g_1=g_2$ case differs from the disjoint case by almost a factor of $3$, while it is practically indistinguishable from the intermediate case.

We summarize our main findings:
\begin{enumerate}
\item Our semianalytic approach merges a wounded constituent model with a string model. We constrained the model to reproduce the one-body spectra in pseudo-rapidity.
\item A family of possible solutions to the string-end-point distributions exists. However they can be further discriminated  (at least for the two limiting cases) via two-particle correlations in rapidity.
\item The Legendre coefficients $a_{nm}$ of the correlations approximately scale as the inverse of the number of sources, as expected.
\end{enumerate}

\bibliographystyle{apsrev4-1}
\bibliography{proc_rohrmoser.bbl}
\end{document}